\documentclass[epj,twocolumn]{webofc}
\usepackage[varg]{txfonts}   
\newcommand{\degr}{^{\circ}}
\newcommand{\etal}{{et al.}}
\woctitle{The Innermost Regions of Relativistic Jets and Their Magnetic Fields}
\begin{document}
\title{Greenland Telescope (GLT) Project}
%
%
\subtitle{''A Direct Confirmation of Black Hole with Submillimeter VLBI''}

\author{M. Nakamura\inst{1}
\and J. C. Algaba\inst{1}
\and K. Asada\inst{1}
\and B. Chen\inst{1}
\and M.-T. Chen\inst{1}
\and J. Han\inst{1}
\and P. H. P. Ho\inst{1}
\and S.-N. Hsieh\inst{1}
\and T. Huang\inst{1}
\and M. Inoue\inst{1}
\and P. Koch\inst{1}
\and C.-Y. Kuo\inst{1}
\and P. Martin-Cocher\inst{1}
\and S. Matsushita\inst{1}
\and Z. Meyer-Zhao\inst{1}
\and H. Nishioka\inst{1}
\and G. Nystrom\inst{2}
\and N. Pradel\inst{1}
\and H.-Y. Pu\inst{1}
\and P. Raffin\inst{1}
\and H.-Y. Shen\inst{1}
\and C.-Y. Tseng\inst{1}
\and the Greenland Telescope Project Team\inst{1,2,3,4}
}

\institute{Institute of Astronomy and Astrophysics, Academia Sinica,
11F of Astronomy-Mathematics Building, AS/NTU No 1, Sec 4, Roosevelt Rd., Taipei 10617, Taiwan
\and
Harvard-Smithsonian Center for Astrophysics, 60 Garden Street, Cambridge, MA 02138, USA
\and
MIT Haystack Observatory, Off Route 40, Westford, MA 01886-1299, USA
\and
NRAO, 520 Edgemont Road, Charlottesville, VA 22903-2475, USA}

\abstract{The GLT project is deploying a new submillimeter (submm)
VLBI station in Greenland. Our primary scientific goal is to image a
shadow of the supermassive black hole (SMBH) of six billion solar masses
in M87 at the center of the Virgo cluster of galaxies. The expected SMBH
shadow size of 40-50 $\mu$as requires superbly high angular resolution,
suggesting that the submm VLBI would be the only way to obtain the
shadow image. The Summit station in Greenland enables us to establish
baselines longer than 9,000 km with ALMA in Chile and SMA in Hawaii as
well as providing a unique $u$--$v$ coverage for imaging M87. Our VLBI
network will achieve a superior angular resolution of about 20 $\mu$as
at 350 GHz, corresponding to $\sim2.5$ times of the Schwarzschild radius
of the supermassive black hole in M87. We have been monitoring the
atmospheric opacity at 230 GHz since August. 2011; we have confirmed the
value on site during the winter season is comparable to the ALMA site
thanks to high altitude of 3,200 m and low temperature of $-50\degr$C. We
will report current status and future plan of the GLT project towards
our expected first light on 2015--2016.}
\maketitle

\section{Introduction}
A direct confirmation of the black hole (BH) in the universe, is one of
the ultimate goal in the modern physics and astronomy. When it is
achieved, we for the first time access matter and electromagnetic fields
under the extremely strong gravity. A BH shadow is expected against the
bright-enhanced annulus of an emission around the BH. The size of
annulus, ``Event Horizon'' as defined by the Schwarzschild radius
($r_{s}$), is lensed and self-magnified by its strong
gravity. Therefore, a detection of the BH shadow is the direct 
confirmation of the existence of BH, and provides a test for General
Relativity under the strong gravity.
In addition to the detection of the strong lensing, the BH spin can be
also probed by a precise imaging of the shape and axis of the shadow,
since it is expected that the BH shadow would be compressed
perpendicular to the spinning axis of the BH. Furthermore, since we
observe BHs as the silhouette against accretion disks and/or
relativistic jets, BH shadow imaging will simultaneously bring us the
images of innermost region of accretion disks and formation region of
relativistic jets as well.


It has been a growing recognition that submillimeter (submm) VLBI
technique could be a unique technique to attain an enough resolution to
resolve nearby SMBHs at the center of AGN, and imaging a shadow of the
BH in the coming decade \cite{Doeleman08}. Since apparent size of
shadows is expected to be very small, practically at this moment, there
are only two possible candidate sources even for submm VLBI: our
galactic center Sgr~A* and one of nearest AGNs M87. The BH mass
$M_{\bullet}$ of Sgr~A* has been estimated by monitoring stellar orbits
around the BH as $M_{\bullet} \simeq 4.3 \times$ 10$^{6}$ M$_{\odot}$
\cite{Gillessen09}. With the current best estimate of the distance of
$8.3$ kpc \cite{Gillessen09} to the galactic center, the BH shadow has
an angular diameter of $\theta = 3\sqrt{3}r_{\rm s} \simeq 52\,\mu$as
($r_{\rm s}\sim 10\,\mu$as). On the other hand, the BH mass is measured
with a range $3.2 \times 10^{9} M_{\odot}$ \cite{Macchetto97} to $6.6
\times 10^{9} M_{\odot}$ \cite{G11} in M87; together with the
distance of 16.7 Mpc to the source \cite{J05}, the largest $M_{\bullet}$
gives the second largest angular diameter of $\theta \simeq 42\,\mu$as
($r_{\rm s}\sim 8\,\mu$as).

Therefore, in order to image a shadow of these SMBHs, an angular
resolution of at least 40--50 $\mu$as in submm VLBI observations would
be required. Very recently, \cite{D12} conducted the Event Horizon
Telescope (EHT) observation at a wavelength of 1.3 mm, deriving the size
of 230 GHz VLBI core to be a FWHM of $40 \pm 1.8\,\mu$as, corresponding
to $5.5 \pm 0.4 \, r_{\rm s}$. This is smaller than the diameter for the
innermost stable circular orbit of a retrograde accretion disk,
suggesting that the M87 jet may be powered by a prograde accretion disk
around a spinning SMBH. Indeed, this work gives a promising insight that
Earth-sized submm VLBI networks are functional to provide angular
resolutions to unveil the SMBHs.

\section{Site Selection}

In July 2010, the US National Science Foundation (NSF) announced a call
for expression of interests for a prototype ALMA 12m telescope, which is
designed from mm to submm wavelength (or 30 to 950 GHz). CfA/ASIAA was
awarded this telescope in April 2011, under collaborating with MIT
Haystack observatory and NRAO to conduct a submm VLBI operation. We have
examined suitable site for allocating a new submm telescope; our main
requirements are (1) excellent atmospheric conditions to perform high
quality observations at submm and even shorter wavelengths, and (2) a
location which provide the longest baselines connecting with other key
stations of submm operations.

Based on the precipitable water vapor (PWV) data mesured by the NASA
satellites, as well as scientific merits and logistics, we have finally
selected Summit Camp in Greenland as the best candidate. Then, M87 has
become our primary target for imaging the BH shadow by the submm VLBI
with other facilities, such as SMA in Hawai, LMT in Mexico, and ALMA in
Chile (Sgr A* cannot be observed from Greenland, as it is on the
southern sky). From here, we name our project as Greenland Telescope
(GLT) project.

\begin{figure}
\centering \includegraphics[width=7.5cm]{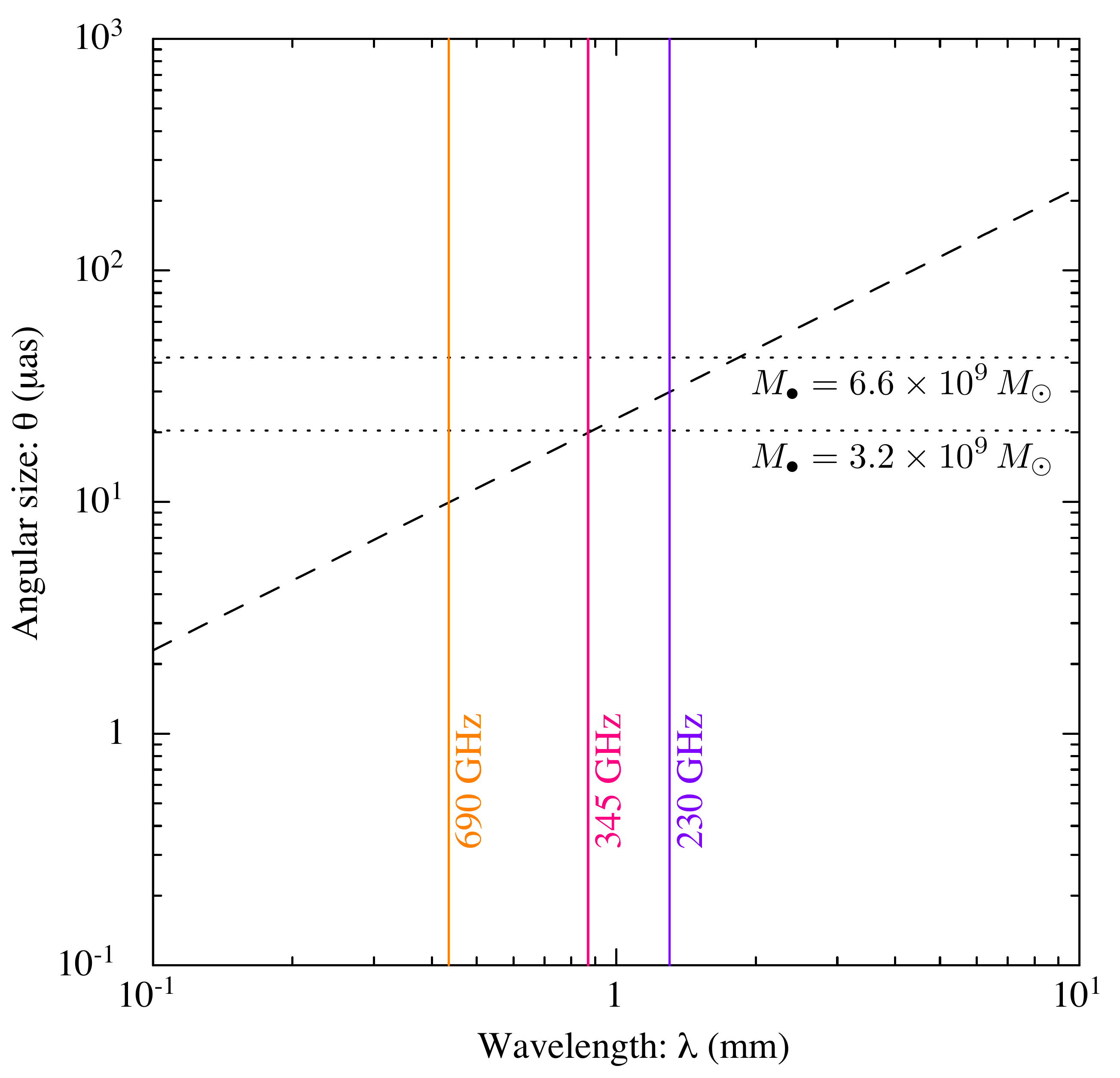} \caption{
Required resolutions to image the black hole shadow in M87. Dashed line:
estimated resolution of a radio interferometer, taken by $\theta_{\rm
res}=k \lambda/d$ with $k=1$ and $d=9,000$ km. Dotted lines: angular
diameter of the BH shadow $\theta=3\sqrt{3}\, r_{\rm s}$, corresponding
the largest and smallest $M_{\bullet}$ cases. Imaging the BH shadow of
M87 will be possible with submm VLBI $\lambda \lesssim 1$ mm, indicated
as vertical solid lines of 230 GHz (purple), 345 GHz (magenta), and 690 GHz
(orange), respectively.}
\label{f0}
\end{figure}

Figure \ref{f0} shows the resolution of a radio interferometer in order
to resolve the BH shadow in M87 at a given baseline $d=9,000$ km, which
is a comparable length between the GLT and the ALMA. This yields
$\theta_{\rm res}\simeq 23\lambda_{\rm mm}$, where $\theta_{\rm res}$ is
the resolution in $\mu$as and $\lambda_{\rm mm}$ is the observed
wavelength in millimeter.  Fig. \ref{f0} indicates submm VLBI
observations at $\lambda \lesssim 1$ mm can fairly resolve the BH shadow
diameter in M87 \cite[see also][for Sgr~A*]{J12}.

\subsection{Atmospheric condition}
We investigated the distribution of the monthly mean of the PWV in 2008
based on data taken by NASA Aqua and Terra/MODIS. It turns out that the
PWV at the inland of the Greenland is less than 2mm though the year,
indicating a promising candidate for submm VLBI observation.

\begin{figure}
\centering \includegraphics[width=8cm]{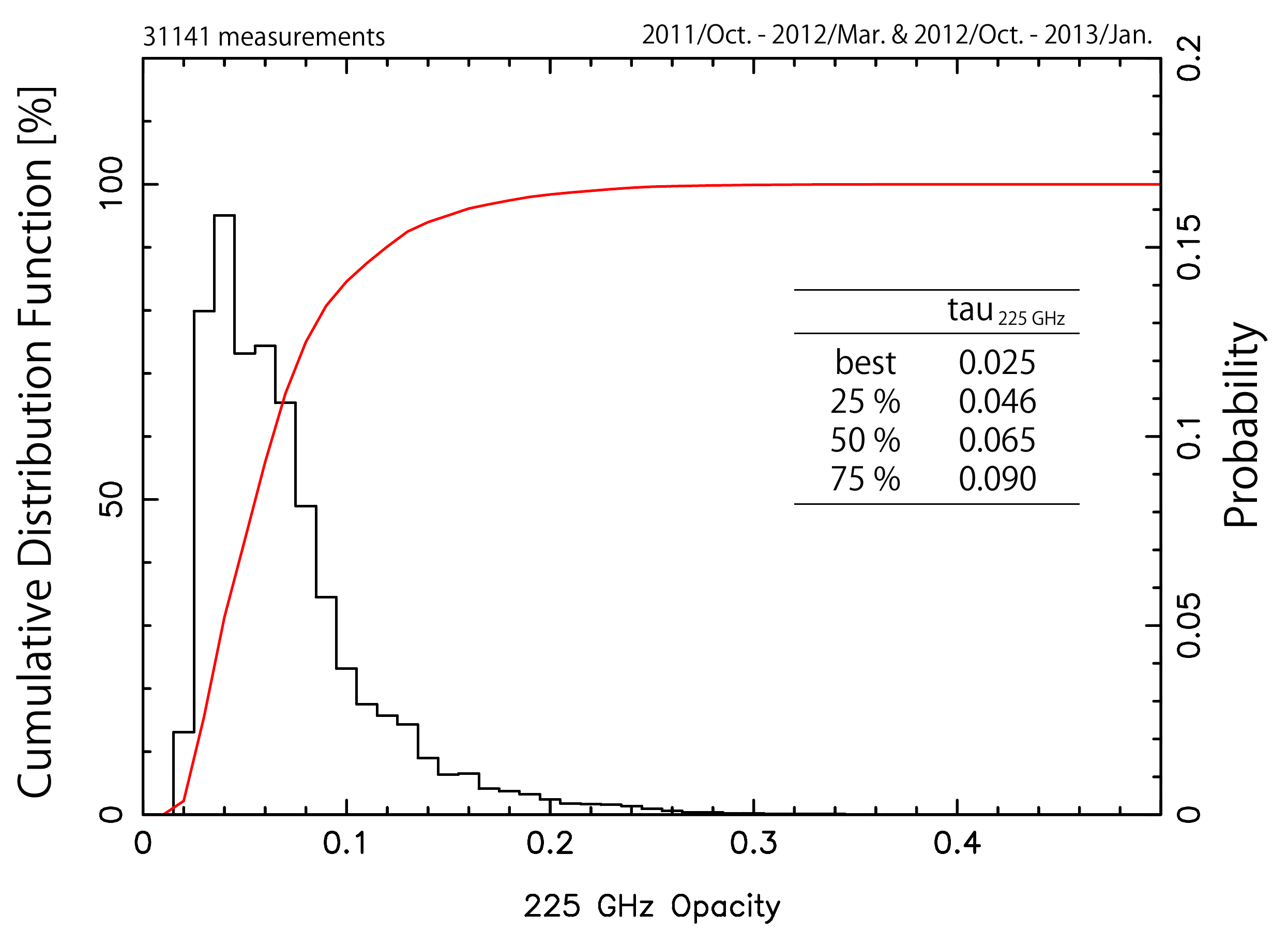}
\caption{
Histogram of the 225 GHz opacity at the Summit Camp, Greenland from 2011
Aug. to 2013 Jan. and its cumulative distribution function.
}
\label{f2}
\end{figure}

We purchased a tipping radiometer from Radiometer Physics GmbH. After a
test run at the summit of Mauna Kea, Hawaii, the radiometer was deployed
to Summit Camp in August 2011. The atmospheric opacity has been
monitored since August 2011 at Summit Camp. The median values of the
measured opacity at Summit Camp varied in the range from 0.04 to 0.18
between August 2011 and Jan. 2013 (Fig. \ref{f2}). Summit Camp on Greenland
is expected to be an excellent site for submillimeter and Terahertz
astronomy.

\subsection{Expected $u$--$v$ coverage }
\begin{figure}
\centering \includegraphics[width=7.5cm]{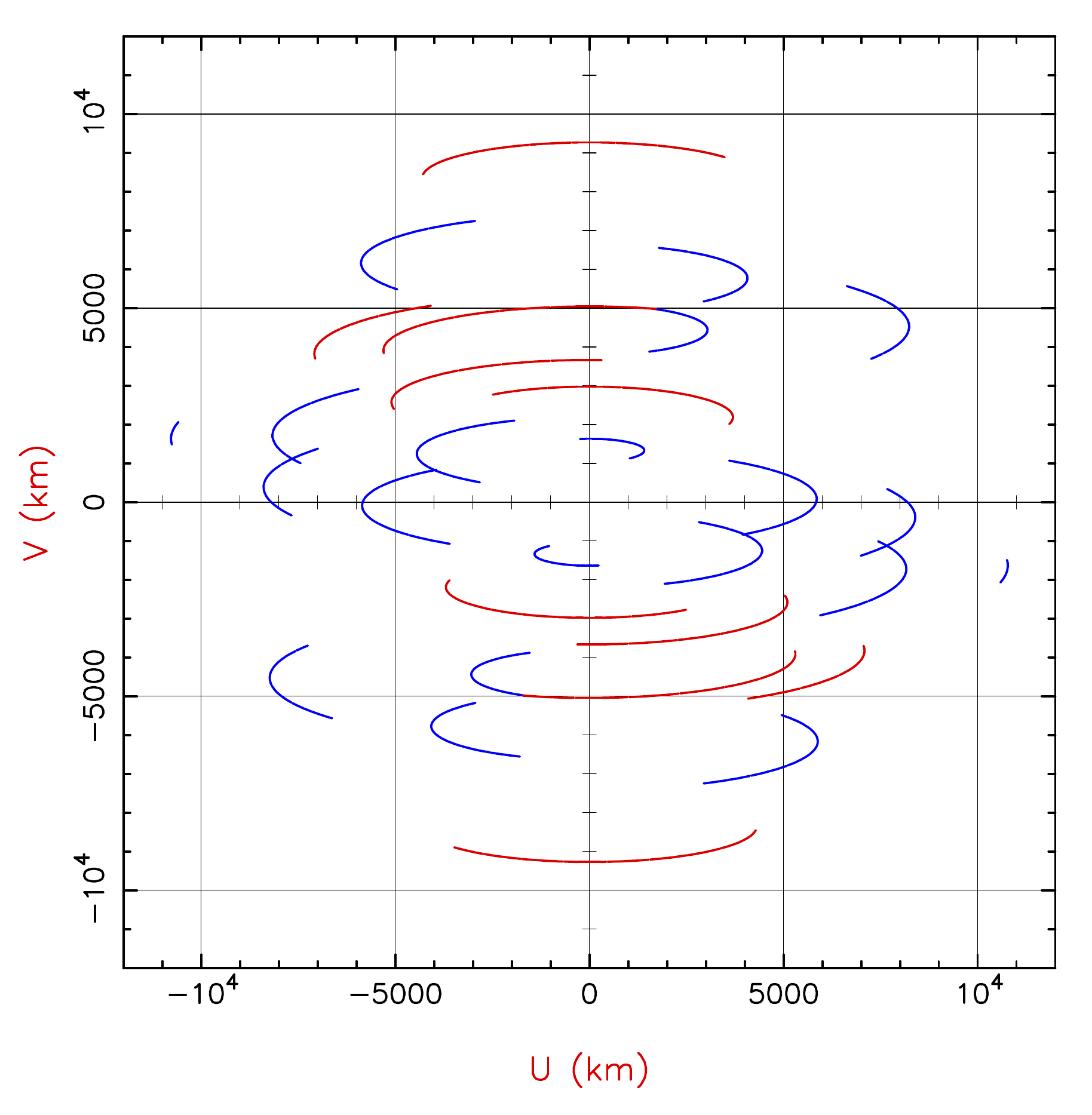}
\caption{
Expected $u$--$v$ coverage towards M87. We assume the frequency at 350 GHz
with telescopes of large aperture size, the ALMA, the SMA, the SMTO in
Arizona, the Large Millimeter Telescope (LMT) in Mexico, and the IRAM
30-m telescope at Pico Veleta in Spain, together with GLT at Summit
Camp. Red lines indicate baselines with the GLT.
}
\label{f3}
\end{figure}

Figure \ref{f3} shows expected $u$--$v$ coverage towards M87 including the GLT
Baselines between the Summit Station and the other stations provide
the longest and unique baselines. The longest baseline is provided by
the combination between the Summit Station and the ALMA, giving the
baseline length of 9,000 km. It provides us an angular resolution of 20
$\mu$as at 345 GHz, which corresponds to half of the expected size of
the BH shadow with $6.6 \times 10^{9} M_{\odot}$ \cite{G11} in M87.

\subsection{Logistics }
The Summit Camp is a geophysical and atmospheric research station,
established and maintained by the U.S. NSF, cooperating with the
Government of Greenland. The Summit Station is located roughly
72.5$^{\circ}$ N, 38.5$^{\circ}$ W (north of the Arctic Circle) with a
3,200 m altitude, and is on top of the Greenland ice cap.  Main purpose
of this station is logistic supports for researchers to conduct
year-round, long-term measurements for monitoring and investigations of
the Arctic environment.  However, NSF funded researchers in any field
can have access to this facility.  The station can be physically reached
by C-130 and small aircrafts. So far, diesel power and a network of 64
kbps VSAT satellite link are available upon request.

\begin{figure}
\centering \includegraphics[width=7.5cm]{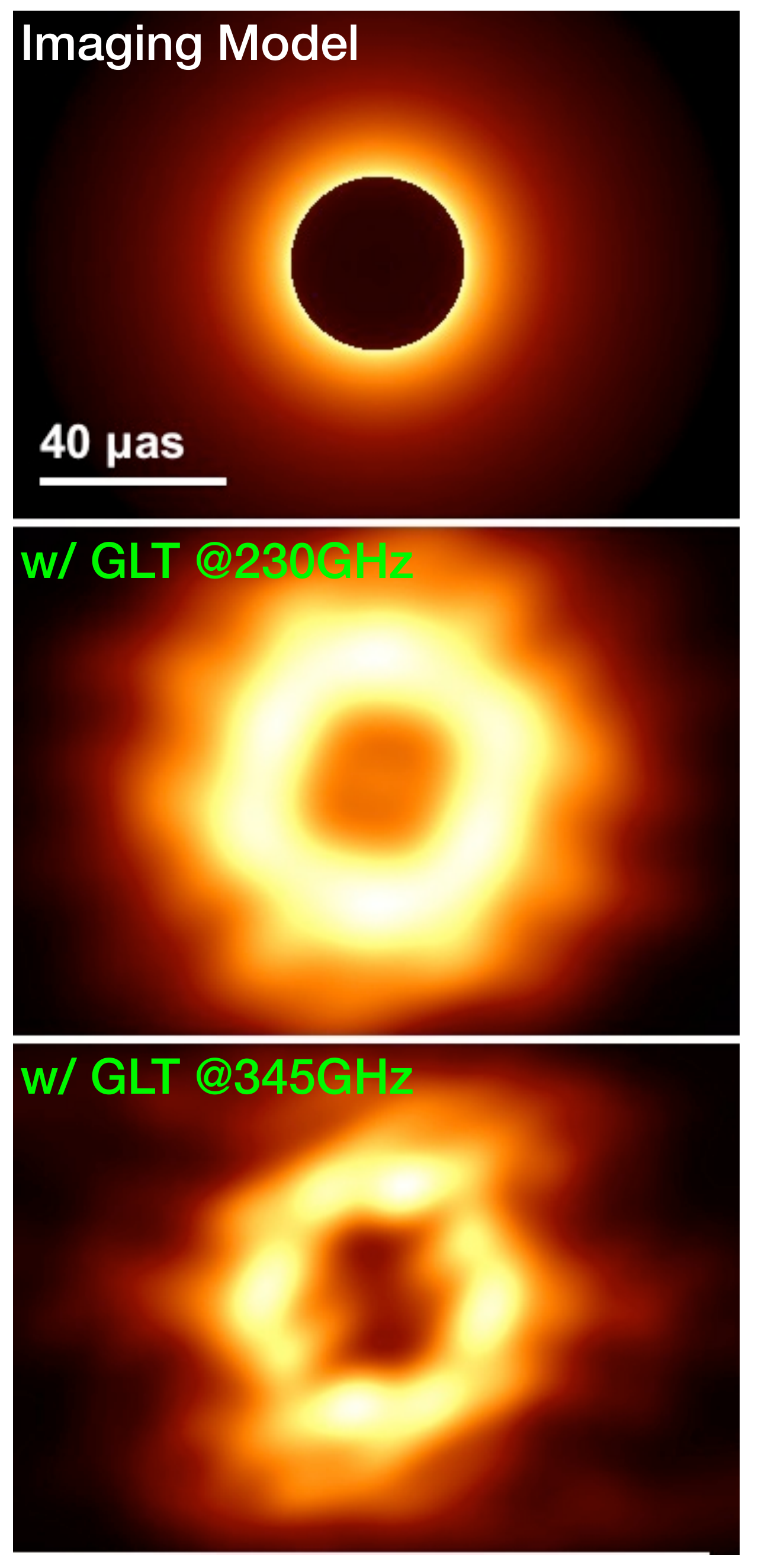}
\caption{
Simulated images of SMBH shadow of M 87. By using the ray-tracing
method, we model the shadow image for the case of a non-rotating, six
billion solar masses SMBH enclosed by optically thin, free-falling
materials. The panel shows the model image (top), and the image obtained
by CLEAN deconvolution algorithm with simulated submm VLBI array
including GLT, of 230 GHz (middle) and 345 GHz (bottom), which are
affected by finite resolution, instrumental thermal noise, and CLEAN
error that deconvolution process introduces.
}
\label{f4}
\end{figure}

\section{Science Case: Imaging Simulation of the SMBH Shadow}

It is thought that (uncharged) SMBHs can be completely specified by two
parameters, their mass and their spin or angular momentum. Although
current methods are able to provide an estimate for both of these, their
extraction is very coarse and typically only very rough limits are
discussed for the spin and very large errors are found for the mass.
The detection of a shadow of nearby SMBHs can provide us with an
alternative, more direct way, to extract their mass and
spin. Furthermore, efforts in order to understand the submm emission on
M87 are currently undergoing and is still unclear if dominant emission
at these wavelengths arises from disk, jet, or a combination of these
\cite{BL09, DMA12}.

A number of simulations is currently undergoing in order to understand
and estimate how these parameters will affect the silhouette of the SMBH
shadow. Inversely, once we are able to obtain an image, we will be able
to constraint the parameters space. Our current models are based on steady
solutions of the radiatively inefficient accretion flows and
ergosphere/disk-driven general relativistic magnetohydrodynamic (GRMHD) jets by (semi-)analytical
formulations. These key ingredients are then used for the ray-tracing and general
relativistic radiative transfer (GRRT) around the SMBH to obtain a
``theoretical'' (infinite resolution) image. We then input this model
into a simulated submm VLBI array including the GLT to understand how
the real observation image is affected by finite resolution, $u$--$v$
coverage, sensitivity and antenna performances \cite{P13}. Figure
\ref{f4} shows one of our imaging samples, suggesting the future submm
observations including the GLT will be capable of resolving the BH
shadow in M87. Note dynamical effects will be taken into account by
implementing the GRMHD simulation and time-dependent GRRT in the future
to examine observed time variability.

\section{Status of the GLT Project}

\begin{figure}
\centering \includegraphics[width=8cm]{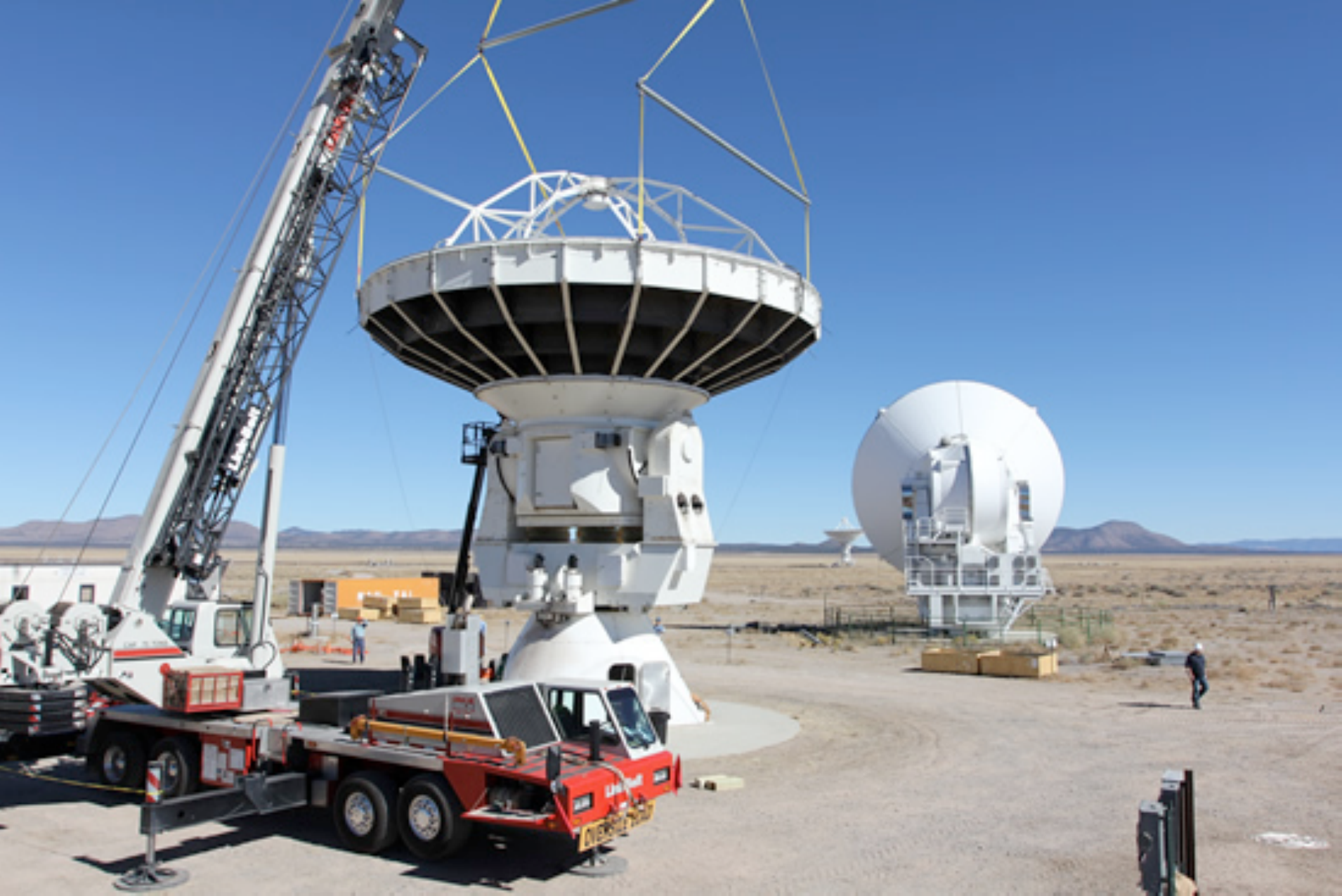} \caption{ Disassemble one
of the ALMA prototype antennas for retrofitting and ready to operate in
the extremely cold environment on the Greenland ice sheet. This picture
shows the disassembly process in November 2012, at the site of the Very
Large Array near Socorro, New Mexico. This effort is led by ASIAA
personnel, Philippe Raffin, Ted Huang, and George Nystrom, in
collaboration with the engineers from Aeronautical Research Laboratory.}
\label{f5}
\end{figure}

\subsection{Antenna}
We plan to use the ALMA-NA prototype antenna (see Fig. \ref{f5}) originally
designed as a prototype for the ALMA. The antenna needs to be
retrofitted to adapt to the new environment condition, since the
environment condition at the Summit Camp is significantly different from
that at the ALMA site.  We have already started to work with Vertex on
it.  We note that Vertex is the original vender of the antenna and has
extensive experience with similar telescope working in similar
environment, the South Pole Telescope.  We have started initial
inspection and functional test from 2011.  We also conducted surface
measurement by photometography and installed a new antenna control
software and an optical pointing telescope before it had been disassembled
for the retrofitting at the NRAO VLA site in New Mexico (see also Fig. 
\ref{f5}).  In November 2012, the antenna was disassembled and shipped to many
sites for retrofitting.

\subsection{Receiver plan}
For the submm VLBI purpose, we are considering to have 86, 230, 345, and
possibly 690 GHz receivers.  For the single dish observation, we are
considering to have heterodyne multi feed receivers and a multi-pixel
imaging array at THz and sub-THz frequency.

\subsection{Future Prospects}
After retrofitting, the antenna will be shipped to Greenland in 2014.
In the meantime, we will work on construction for the foundation and
infrastructures with help of NSF US and CH2MHILL Polar Services.  We
hope to have the first light in 2015/16.

\section{VLBI Data Acquision}

\subsection{ALMA Phase-up project}
The introduction of ALMA antennas will be a key in the submm VLBI
network, as it will be able to increase the sensitivity by a factor of
ten due to its large collecting area. ASIAA has a very active
participation in an international collaboration led by MIT Haystack
Observatory aiming towards the ALMA phase-up project.  An international
consortium is presently constructing a beamformer for the ALMA in Chile
that will be available as a facility instrument. The ALMA beamformer
will have impact on a variety of scientific topics, including accretion
and outflow processes around black holes in active galactic nuclei
(AGN), tests of general relativity near black holes, jet launch and
collimation from AGN and microquasars, pulsar and magnetar emission
processes, the chemical history of the universe and the evolution of
fundamental constants across cosmic time, maser science, and astrometry \cite{F13}.

\subsection{DiFX correlator}
ASIAA has acquired a CPU cluster with broadband interconnect network
connection (infini-band) capabilities and started the development of a
DiFX software correlator, which is easy to maintain and upgrade and will
be able to handle the massive data rate from submm VLBI observations.
We have correlated test data and obtain fringes using 1.3 mm VLBI data
from the EHT observation in 2012. We aim to solve discrepant sampling
schemes between ALMA and other VLBI stations, including GLT, by DiFX
enhancements.



\end{document}